\documentclass[prl,twocolumn,amsfonts,showpacs]{revtex4-1}

\pdfoutput=1

\usepackage{epsfig}
\usepackage{psfrag}
\usepackage{amsmath}
\usepackage{amssymb}
\usepackage{color}
\usepackage{hyperref}

\begin{document}
\author{Chuan-Yin Xia}
\author{Hua-Bi Zeng}
\email{hbzeng@yzu.edu.cn}

\affiliation{ Center for Gravitation and Cosmology, College of Physical Science
and Technology, Yangzhou University, Yangzhou 225009, China}

\title{ Kibble Zurek mechanism in  rapidly  quenched  phase transition dynamics}

\begin{abstract}
We propose a  theory  to explain the experimental observed
deviation from the  Kibble-Zurek mechanism (KZM)  scaling 
in  rapidly quenched critical phase transition dynamics.
There is a critical quench rate $\tau_{Q}^{c1}$ above it the KZM scaling begins to  appear.
Smaller than  $\tau_Q^{c1}$, the defect density $n$ is a constant independent of the
quench rate but depends on the final temperature $T_f$ as $n \propto L^d \epsilon_{T_f} ^{d \nu}$,
the freeze out time $\hat{t}$ admits the scaling law $\hat{t} \propto \epsilon_{T_f}^{-\nu z}$
where $d$ is the spatial dimension, $\epsilon_{T_f}= (1-T_f/T_c)$ is the dimensionless reduced  temperature, $L$ is the sample size, $\nu$ and $z$ are spatial and dynamical critical exponents.
Quench from $T_c$, the critical rate is determined by the final temperature $T_f$
as $\tau_Q^{c1} \propto \epsilon_{T_f}^{-(1+z \nu)} $. All the scaling laws are verified in a rapidly quenched superconducting ring
via the AdS/CFT correspondence.
\end{abstract}
\maketitle
\begin{figure}[t]
\centering
\includegraphics[trim=1.5cm 10.4cm 5.9cm 9.9cm, clip=true, scale=0.6, angle=0]{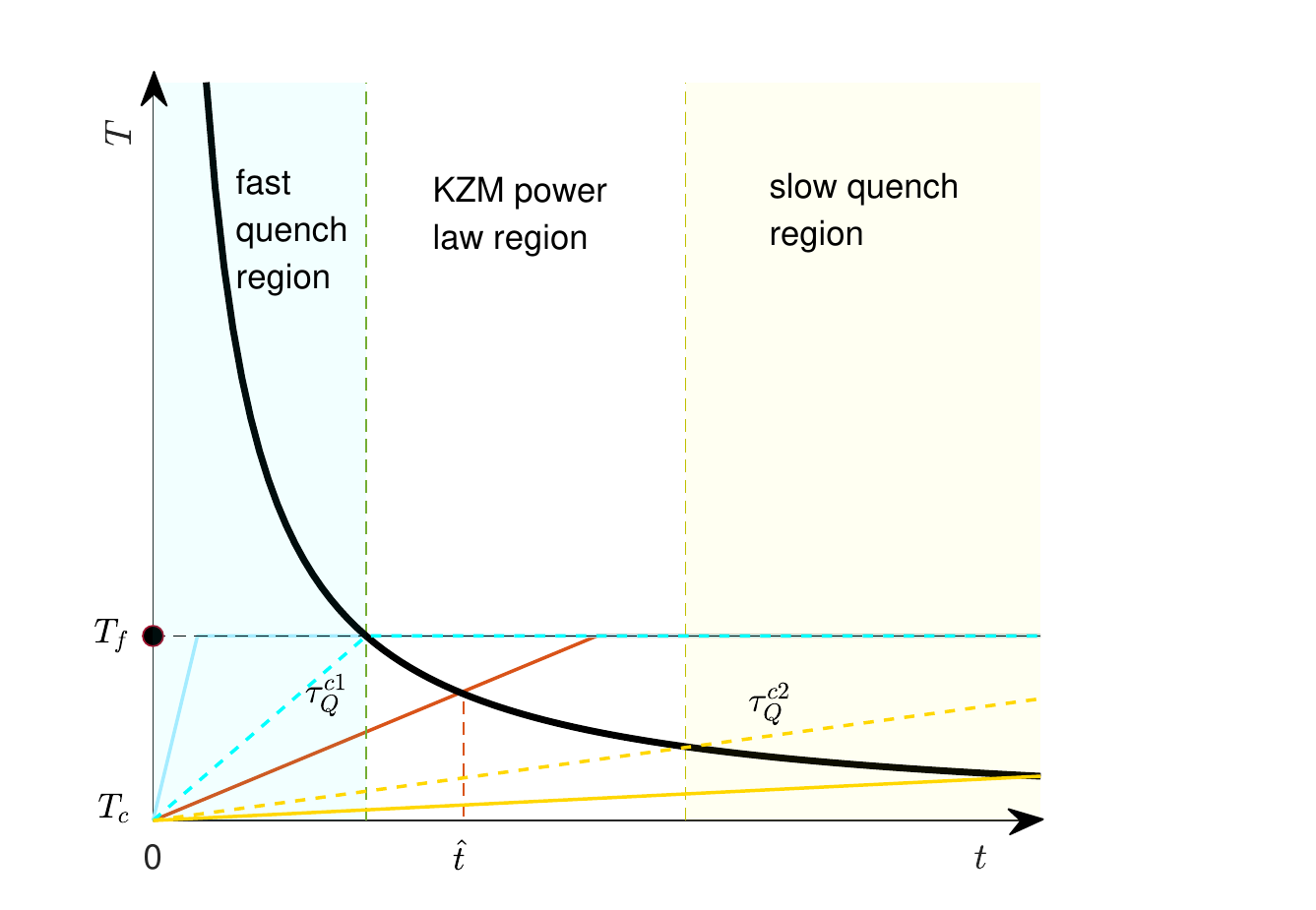}
\caption{Schematic representation of the freeze-out captured by the adiabatic-impulse
approximation for different quench rates. The black line is the equilibrium relaxation time $\tau(T)$,
the other lines denote the piece-wise function for the quenched temperature $T(t)$.
In the KZM power law region, the $T(t)$ line intersects the $\tau(T)$ line before the end of quench, while in the
fast quench region the $T(t)$ lines have the same intersection point with the  $\tau(T)$ line at the time $\tau(T_f)$ }\label{Fig1}
\end{figure}

\textit{Motivation: }The Kibble-Zurek mechanism (KZM) \cite{Kibble1976,Kibble1980,Zurek1985,Zurek1996,Polkovnikoc2011,Campo2014}.
is one of the cornerstones  of out-of equilibrium condensed matter physics,
which gives a universal framework to understand the universal scaling law in  phase transition dynamics.
The schematic picture of KZM is based on the the adiabatic-impulse
approximation, which states  that when a system crosses the
critical temperature to the low temperature phase
by a linear quench, the divergent equilibrium  relaxation time at critical point
prevents the system to evolve promptly until the  time after quench is equal to the equilibrium
relaxation time $\tau[T(\hat{t})]$, the red line in  Fig. \ref{Fig1}.
The time $\hat{t}$ is defined as the freeze out time at the moment the system
catches the quench speed then begin to evolve quickly,
which can be calculated by solving the equation
\begin{equation}
\hat t=\tau[T(\hat{t})].
\label{Eq1}
\end{equation}
With  the universal scaling law of relaxation tim
$\tau(T) \sim \tau_0 \epsilon ^{-\nu z}$
near $T_c$ where $\epsilon= (1-T/T_c)$, the freeze time can be obtained as
\begin{equation}
\hat{t} \sim (\tau_0 \tau_{Q}^{z \nu})^{\frac{1}{1+z \nu}},
\label{Eq2}
\end{equation}
where the linear quench rate $\tau_Q$ is defined from the
formula of quenched temperature $T(t)= T_c(1-\frac{t}{\tau_Q})$, the $t/\tau_Q$
is the dimensionless
distance from the critical point.
At the moment the order parameter begins to develop dramatically,
independent spatial domains will form randomly, whose  average size is  expected to be the equilibrium  correlation length of the order parameter
at $T(\hat{t})$, then topological defects will be generated at the interfaces of the domains. The defect density $n$ is proportional
to the number of domains $L^d/\xi(T(\hat t))^d$. From Eq. \ref{Eq2} the reduced temperature at freeze out time $\hat{t}$ is decided by quench rate $\tau_Q$
as $(1-T(\hat{t})/T_c) \sim (\frac{\tau}{\tau_Q})^{\frac{1}{1+z \nu}}$, plug the temperature to the correlation length formula
$\xi(T) \sim \xi_0 \epsilon^{-\nu} $ near $T_c$ the average defect density $n$ formed in the dynamic process can be estimated
\begin{equation}
n \sim \frac{L^d}{\xi(\hat{t})^d} \propto \tau_{Q}^{-\frac{d \nu}{1+z\nu}}.
\label{Eq3}
\end{equation}

The universal scaling law  between defect density and quench rate is the key prediction of KZM, which has been experimentally
tested by using many controllable systems, such as liquid crystals \cite{Chuang1991,Bowick1994,Digal1999},
superfluid helium \cite{Hendry1994,Bauerle1996,Ruutu1996,Dodd1998}, Josephson junctions \cite{Carmi2000,Monaco2002,Monaco2003,Monaco2006}, thin film superconductors\cite{Maniv2003,Golubchik2010},  linear optical
quantum simulator\cite{Xu2014}, trapped ions\cite{Pyka2013,Ulm2013,Ejtemaee2013} and Bose Einstein condensate \cite{Sadler2006,Weiler2008,Lamporesi2013,Navon2015,Donadello2016,Ko2019}.
Notice that the density $n$ in Eq. \ref{Eq3} is a statistical mean quantity,
theoretically, it was found for the first time in \cite{Campo2018,Ruiz2020} that the
statistic distribution of $n$ admits an universal statistical law with with it's average value unchanged  for a given quench rate, which is
crucial for the KZM  prediction.
The universal statistical law has  been observed in a 2D quenched holographic
superconductor\cite{Campo2021}. Experimentally, it
has been verified by a 1D quantum simulation \cite{Cui2020} and also in a rapidly quenched 2D Bose gas \cite{Goo2021}.

Besides the celebrated verification of  KZM scaling, the deviations from KZM prediction
were also reported in rapidly quenched two spatial dimensional atom gas when it undergoes a normal-to-superfluid phase transition ,
where the defect density formatted after the quench demonstrates a plateau whose value is
a constant independent of the quench rate \cite{Donadello2016,Ko2019,Goo2021}.
Numerical simulation also found the same phenomenon by quenching the external potential of an ion chains
through the critical point \cite{Campo2010}.
A plausible explanation of the plateau is attributed to the fast relaxation of defects via pair annihilation
at high density,  
then the experiments are incapable to
detect individual vortices in the  turbulent condensate after fast quench \cite{Donadello2016,Ko2019,Campo2010,Liu2018v1}.
Also, a theoretical study  found that when the quench time is shorter than the
condensation growth time, the defect formation dynamics are similar, which
results in the constant defect density independent of quench rate \cite{Chesler2015}.
However, both explanations are far from complete, there are questions  still need to be answered:
(a) what is the exactly mechanism why the plateau appears? (b) where the plateau ends?  (c) how the defects density depends on the quench ending temperature?
In this letter we give a satisfied scheme to  clarify all the above issues.

\begin{figure}[t]
\centering
\includegraphics[trim=1.5cm 10.4cm 4.9cm 9.9cm, clip=true, scale=0.6, angle=0]{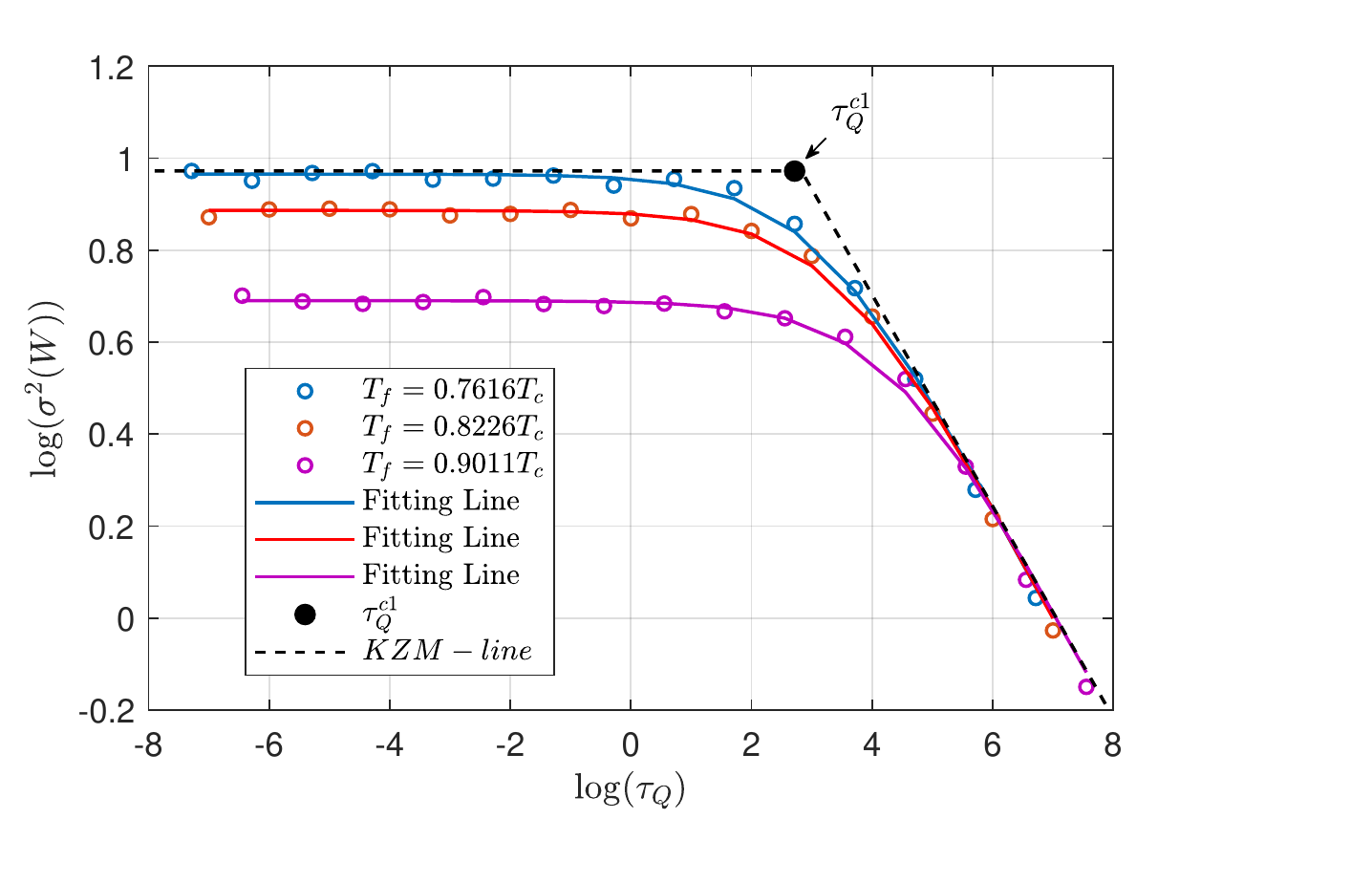}
\caption{Universal scalings of invariant of winding number  versus quench rate in the
quenched 1D holographic superconducting ring for three different quench final temperature.
The black dotted line is from KZM scheme, the Eq. \ref{Eq9} is used to fit the data in the whole quench region. }\label{Fig2}
\end{figure}

\textit{
Improved  KZM under fast quench:} The key observation of KZM is to identify  the moment $\hat{t}$
where the system begin to partition into independent
domains, whose average size is set by the equilibrium coherence length
of the temperature at the moment.  The solution Eq. \ref{Eq2} of $\hat{t}$
only valid when the $T(t)$ line has a intersection point with the $\tau(T)$ line
before the ending of quench  (the red line in Fig. \ref{Fig1}).
When the quench is implemented rapidly, the quench ends before
the time that the $T(t)$ line meets the $\tau(T)$ line (the solid blue line in Fig. \ref{Fig1}). In this cases,
the system will admit the same freeze out time $\hat{t}=\tau(T_f)$ since all quenched temperature line have the
same intersection point  with the $\tau(T)$ line, independent of the values of $\tau_Q$. As a result, the
average sizes of KZM independent domains for different quench rates will have same value $\xi(T_f)$,
which naturally explains the plateau of defect density  appears  in rapidly quenched dynamics.
Furthermore, we can predicted the relationship between defects density and the final
temperature in the fast quench plateau region
\begin{equation}
n \sim \frac{L^d}{\xi(T_f)^d} \propto L^d \epsilon_{T_f}^{d \nu}.
\label{Eq4}
\end{equation}
This is the first prediction, the second prediction of freeze out time is more straight forward,
the $\hat t$  just equals to the relaxation time at $T_f$
\begin{equation}
\hat{t} \sim \tau(T_f) \propto \epsilon_{T_f}^{-\nu z}.
\label{Eq5}
\end{equation}
The first critical quench rate $\tau_Q^{c1}$  is defined when
the quench ending time $\tau_Q^{c1} (T_c-T_f)/T_c$ equal to the relaxation time at $T_f$ as
shown  by the blue dotted line in Fig.\ref{Fig1}
\begin{equation}
\frac{\tau_Q^{c1} (T_c-T_f)}{T_c}=\tau_0 (T_c-T_f)^{-\nu z},
\label{Eq6}
\end{equation}
 solving the equation we have
\begin{equation}
\tau_Q^{c1} \propto   \epsilon_{T_f}^{-(\nu z+1)},
\label{Eq7}
\end{equation}
this is the third prediction.
According to the picture, the defects density is a constant when the quench rate
is lower than $\tau_Q^{c1}$, then  turning to the KZM power law line in an un-smooth way,
as shown by the black dotted line in Fig. \ref{Fig2}. So  The critical quench rate $\tau_Q^{c1}$ can also be defined as  the intersection
point of the two lines. By equal Eq. \ref{Eq3} and Eq. \ref{Eq4} the intersection point can
be computed, which is exactly Eq. \ref{Eq7}.
Until now we clarified the mechanism why the deviation from KZM prediction appears in the
rapidly quenched critical phase transition, three scaling laws are predicted,
which will be verified in a quenched one dimensional superconducting ring.

\textit{Numerical experiment in a rapidly quenched superconducting ring:}
According to KZM analysis \cite{Das2012,Sonner2015}, quench a 1D superconducting/superfluid ring with circumference $C$ will spontaneously generate winding number $W=\oint d \theta/2\pi $,
whose variance $\sigma^2(W)$ admits the KZM scaling $\sigma^2(W) \sim \frac{C}{\xi(\hat{t})} \propto \tau_Q^{-\nu/(1+z \nu)}=\tau_Q^{-1/4}$ when the quench rates are in a proper region and the system admits the mean field critical exponents  that $z=2$ and $\nu=1/2$.
The scaling law of $\sigma^2(W)$ has been confirmed in an experiment of  Bose gas \cite{Corman2014}, and also by theoretical simulation
via the Gross-Pitaevskii  equation \cite{Das2012}, or the holographic superconductor model simulation \cite{Sonner2015,Xia2020}.

\begin{figure}[t]
\centering
\includegraphics[trim=1.5cm 11.4cm 3.1cm 9.9cm, clip=true, scale=0.5, angle=0]{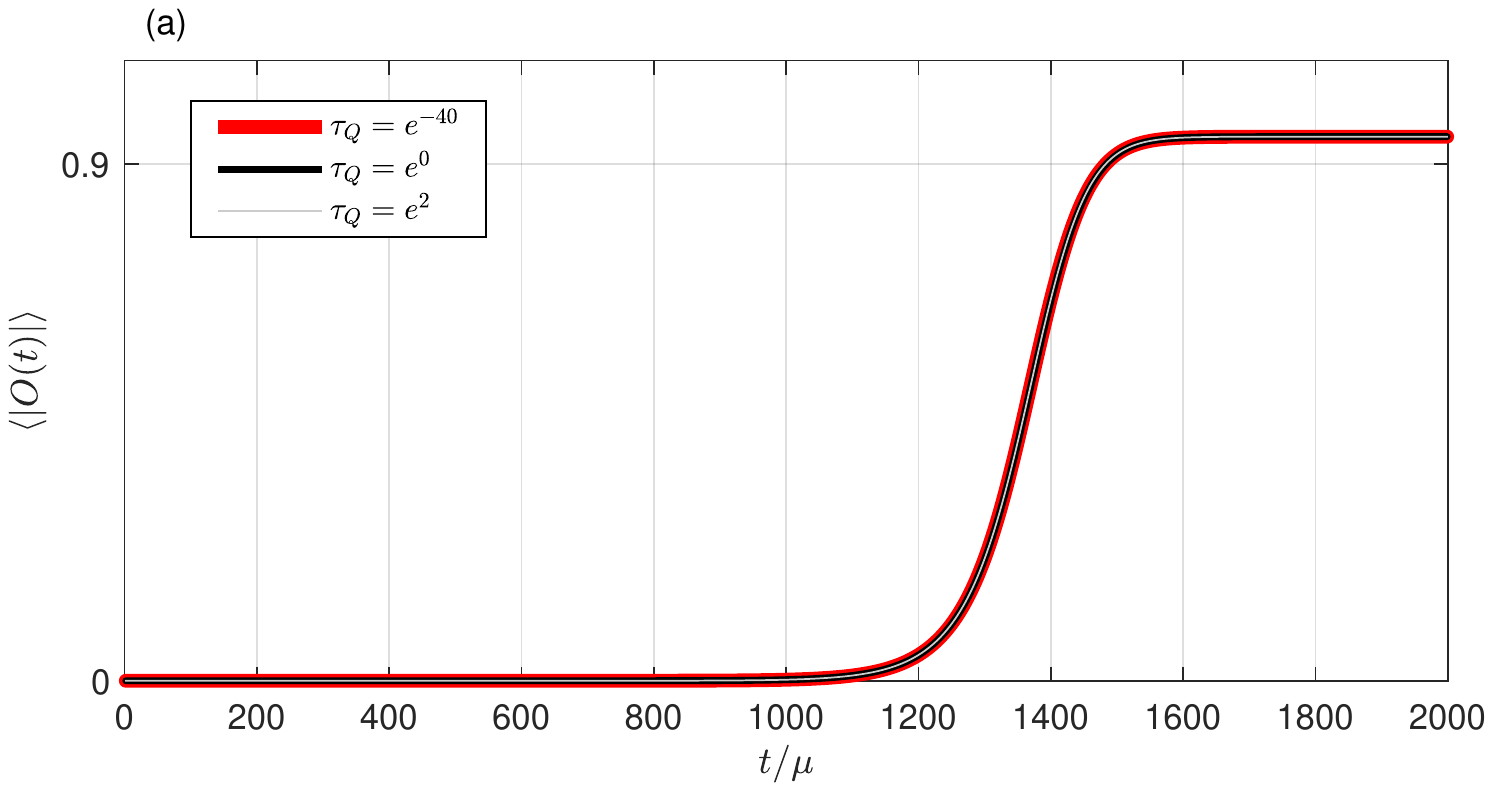}
\includegraphics[trim=1.3cm 10.6cm 2.4cm 10.4cm, clip=true, scale=0.5, angle=0]{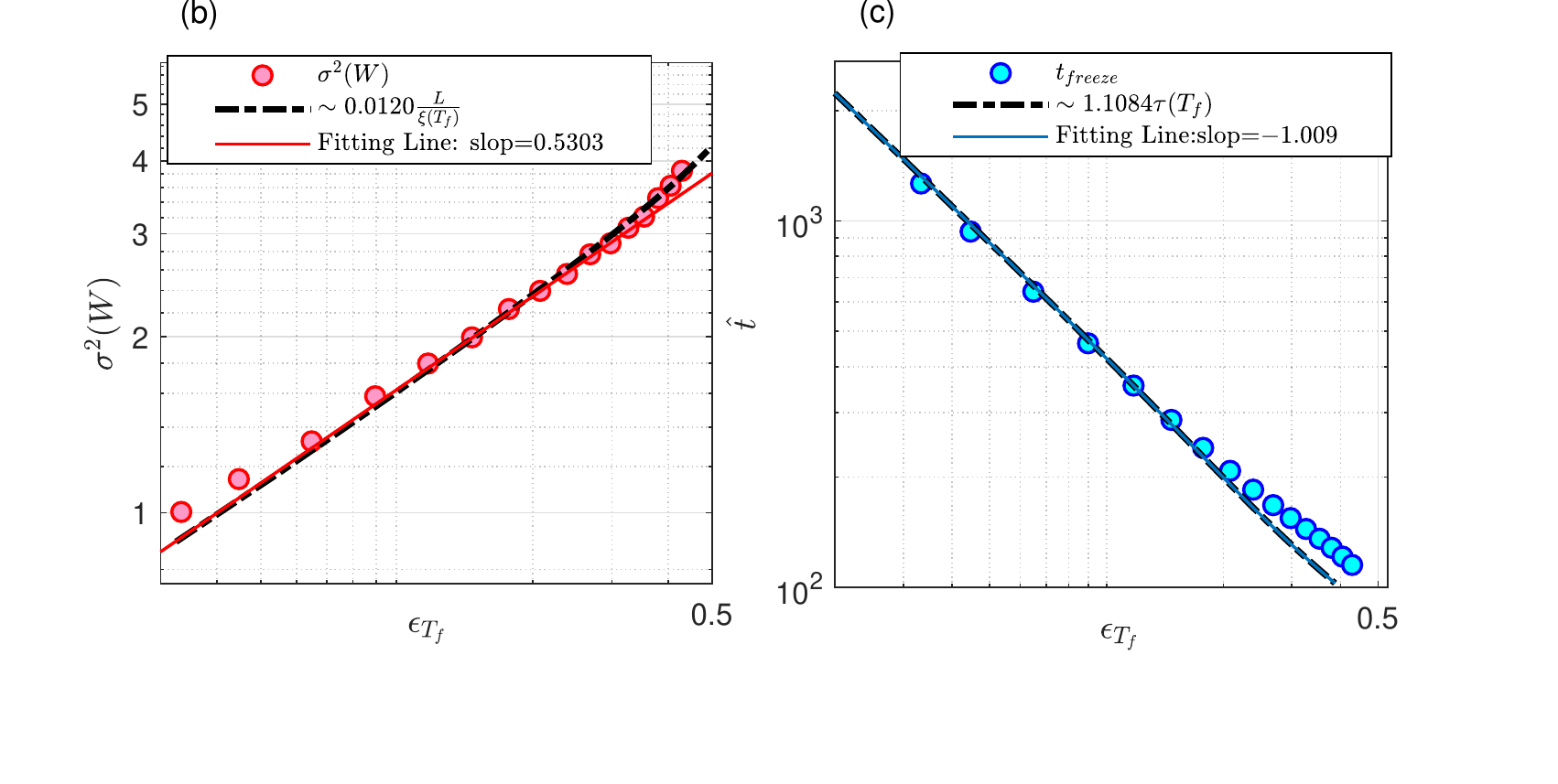}
\caption{(a) Dynamics of average absolute value of the order parameter for three sample different quench rates in the fast
quench region when $T_f=0.9011 T_c$. Before $\hat{t}$, the order parameter
will cease to follow or even approximate its equilibrium value, $\hat{t}$ is defined as the
moment when the average absolute value reaches $0.01 \langle|O_f|\rangle$, $O_f$ is the
equilibrium value of the quench ending temperature $T_f$.
 (b) Invariance of winding number $W$ versus the reduced quench ending temperature $ \epsilon_{T_f}$.  (c) Freeze out time versus $\epsilon_{T_f}$.}\label{Fig3}
\end{figure}

Holographic duality
equates certain strongly correlated systems of quantum
matter without gravity to classical gravitational systems
in a curved space-time with one additional radial spatial dimension \cite{Maldacena1998,Gubser1998,Witten1998},
which provides an  efficient new scheme to study quantum many body systems \cite{Zaanen:2015oix,Ammon2015,Liu2018,Jan2021}.
We adopt a well studied holographic superconducting model developed from AdS/CFT correspondence,
with it's ending temperature can be  tuned conveniently \cite{Gubser2008,Hartnoll2008}.
In the model the boundary superconductor is dual to a classic gravity theory in one more
spatial curved space time, an $AdS$ black hole is coupled with charged scalar  field and $U(1)$ gauge
field
\begin{equation}
S=\int  d^4x \sqrt{-g}\Big[-\frac{1}{4}F^2-( |D\Psi|^2-m^2|\Psi|^2)\Big].
\label{Eq8}
\end{equation}
The  Einstein-Maxwell-complex scalar model defined in the black hole background geometry  $ds^2=\frac{\ell^2}{z^2}\left(-f(z)dt^2-2dtdz + dx^2+ dy^2\right)$ in the
Eddington coordinate, $f(z)=1-(z/z_h)^3$, the black hole temperature $T=3/(4 \pi z_h)$. $x,y$ are the boundary spatial coordinates.
There is a critical value of the black hole temperature below that the charged
scalar develops a finite value in the bulk while it's dual field theory operator have a finite expectation value
$\langle O \rangle$, which breaks the $U(1)$ symmetry in the boundary field theory.
Working in 1D spatial boundary geometry by only turning on coordinate $x$ dependence of all the
fields in the equations of motion, using the periodic boundary condition  in the $x$ coordinate we are effectively studying a superconducting
ring. By solving the dynamic equations by decreasing the black hole temperature
cross $T_c$ the quench induced winding number formation process
can be monitored, then the KZM scaling $\sigma^2(W) \propto \tau_Q^{-1/4}$
was reproduced for a fixed quench ending temperature $T_f=0.8226 T_c$ \cite{Xia2020}.

Here we extend our  simulation \cite{Xia2020} by quenching  the  superconducting ring of circumstance $C=150$  from $T_c$ to different temperatures below $T_c$.
In Fig. \ref{Fig2} sample plots between $\sigma^2(W)$ and quench rates are given for three different $T_f$.
The dashed black line is the KZM predicted line, which matches the numerical simulation, except in the
region near $\tau_Q^{c1}$. We find that the continuous crossover from the
plateau line to the KZM scaling line near $\tau_Q^{c1}$ is similar the experimental observation in \cite{Goo2021}, which can be fitted very well by the following
formula
\begin{equation}
\sigma^2(W)= \sigma_{T_f}^2(W)(1+\frac{\tau_Q}{K(T_f)})^{-\frac{1}{4}},
\label{Eq9}
\end{equation}
Where the coefficient $K(0.9011 T_c)=78.2558$, $K(0.98226 T_c)=32.5014$, $K(0.7616 T_c)=23.2544$.

Also in Fig \ref{Fig1} the KZM scaling Eq. \ref{Eq3} is confirmed when the quench rate is in a proper intermediate region,
where the $\sigma^2(W)$ depends only on $\tau_Q^{c1}$ rather than the
quench ending temperature $T_f$, agrees with the KZM. In the fast quench
region, the $\sigma^2(W)$ demonstrates a plateau whose value depends on $T_f$.
In Fig. \ref{Fig3} (a) it is shown that  average absolute value of the order parameter for
different quench rates admit exactly the same dynamics,
matches our prediction based on KZM that in the fast quench region the system evolves very
similarly.
Furthermore, the prediction  $n \propto \frac{1}{\xi(T_f)} \propto \epsilon_{T_f}^{0.5303} $ is confirmed as plotted in Fig. \ref{Fig3} (b), also the freeze out time agrees
with the prediction  $\hat{t} \sim \tau(T_f) \propto \epsilon_{T_f}^{-1.009}$ as shown in Fig. \ref{Fig3} (c). Here  the holographic superconducting ring is of the mean field equilibrium critical exponents, the value of correlation length and relaxation time are
taken from  \cite{Zeng2021}. The critical quench rate $\tau_Q^{c1}$ where the KZM scaling
begins is defined as  the intersection point of  the plateau line  and the KZM power law line, the prediction of $\tau_Q^{c1}$ in Eq. \ref{Eq7} can be derived by equal the already confirmed  Eq. \ref{Eq3} and Eq. \ref{Eq4}. Then we verified all the three predictions
in the holographic 1D superconducting ring.

\begin{figure}[t]
\centering
\includegraphics[trim=3.5cm 10.4cm 4.9cm 9.9cm, clip=true, scale=0.6, angle=0]{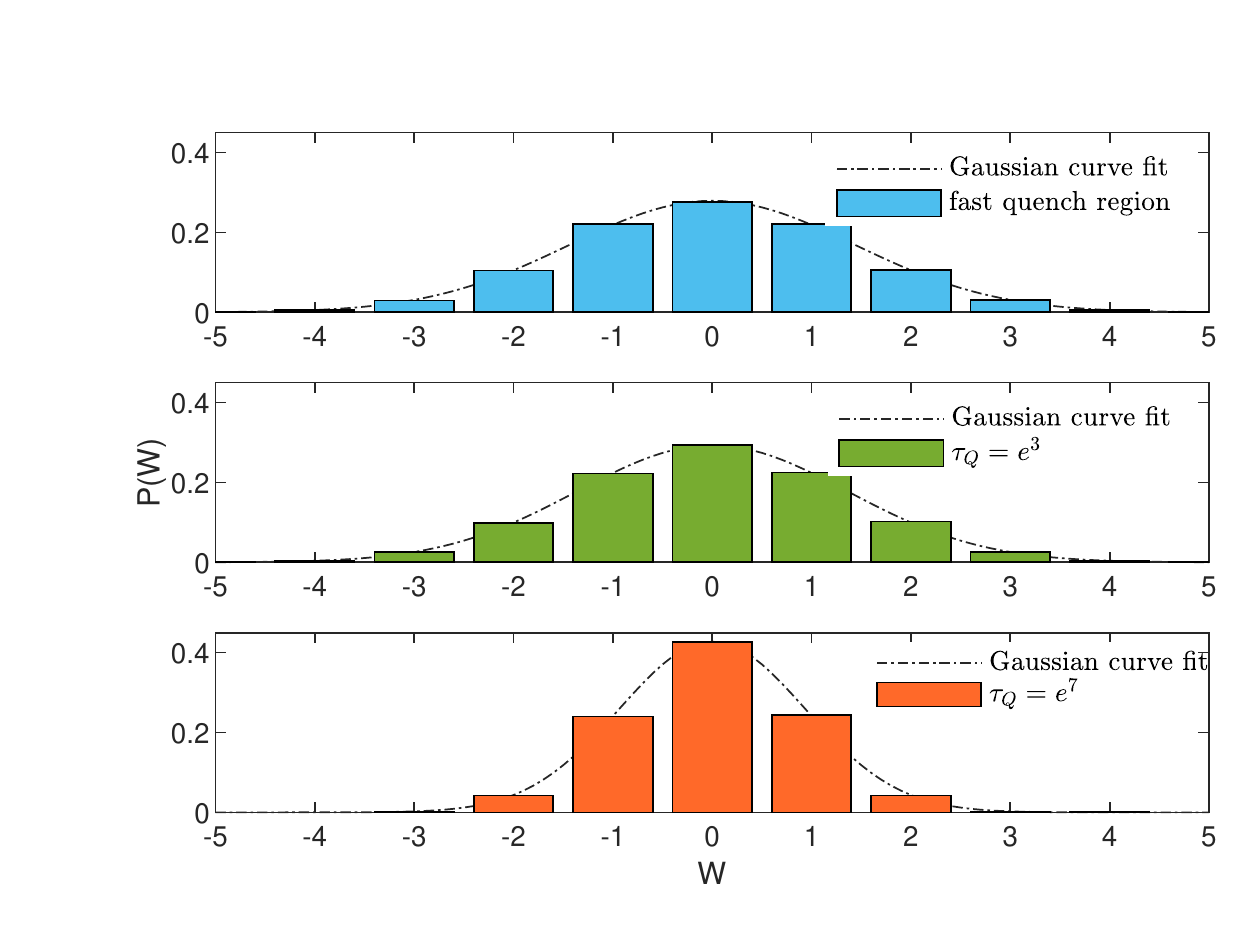}
\caption{Statistic of winding number $W$ for different quench rates when $T_f=0.8226 T_c$.
Each distribution is
obtained from a collection of $5 \times 10^4$ measurements for a given $\tau_Q$. The solid lines denote Gaussian curves fit to the data. }\label{Fig4}
\end{figure}

In Fig. \ref{Fig4} the statistics of winding number $W$ is plotted for different quench rates,
include both fast quench region and KZM power law region.
They all admit the same Gaussian distribution,
which  supports our theory that the defect formation dynamics
in the fast quench region are still of  the KZM scheme style.
Also, the observation eliminate the vortex relaxation scenario, 
matches the experiment observation in a fast quenched Bose gas \cite{Goo2021}.

\textit{Discussion:} We extend the KZM scheme to the fast quench region  to
explain the deviation from the KZM's key scaling law observed by experiments,
the defect density, freeze out time and  the critical quench rate are found to scale with the quench ending
reduced temperature $\epsilon_{T_f}$ in universal power laws, whose powers are  functions of  equilibrium critical exponents $\nu$ and $z$.
The predictions has been confirmed numerically via holography by quenching a superconducting ring.
The  scaling laws are also  accessible to experiments, a
potential candidate is the thin film superconductor whose quench ending temperature can be tuned \cite{Maniv2003,Golubchik2010}.
Though we are dealing with a thermal phase transition induced by temperature, the scaling laws predicted should also
hold in a quenched quantum phase transition  controlled by a
reduced parameter away from the critical point \cite{Ruiz2020}.
We also find the KZM scheme fails to explain the suppress  of defect density and   smooth crossover from
plateau to power law near $\tau_Q^{c1}$, though an
good fitting line  Eq \ref{Eq9}  is proposed but the physics behind is still needed to be revealed,
a reasonable conjecture  is that when the quench ending time
close to the relaxation time at $T_f$, there is a kind of resonance phenomena where the
KZM independent domains form in a  periodic of time rather a particular time point.

  Finally we turn to the case when the quench rate is sufficient large, the yellow line in Fig. \ref{Fig1}, the intersection point is close to
$T_c$ then results a large value for average independent domains  size.
In this case the domain size is expected to be close to the system size, an experiment on a quenched Bose gas ring \cite{Corman2014} found that  the finite size effect will dominate and suppress the defect formation, then  a decay of density with a larger power than KZM prediction  appears.
By quenching a holographic superconducting ring, we reproduced  the
experiment observed deviation from KZM scaling when the correlation length at $\hat{t}$ is larger than a
critical value \cite{Xia2020}.
Due to the monotone increasing relation between quench rate and the $\xi(T(\hat{t}))$, there is a critical quench rate $\tau_Q^{c2}$  when the finite size effect becomes dominate,  larger than it is the slow quench region
as illustrated in Fig. \ref{Fig1}.

We thank Adolfo del Campo for valuable comments.
This work is supported by the National Natural
Science Foundation of China (under Grant No. 11675140).



\begin{thebibliography}{0}%
\makeatletter
\providecommand \@ifxundefined [1]{%
 \@ifx{#1\undefined}
}%
\providecommand \@ifnum [1]{%
 \ifnum #1\expandafter \@firstoftwo
 \else \expandafter \@secondoftwo
 \fi
}%
\providecommand \@ifx [1]{%
 \ifx #1\expandafter \@firstoftwo
 \else \expandafter \@secondoftwo
 \fi
}%
\providecommand \natexlab [1]{#1}%
\providecommand \enquote  [1]{``#1''}%
\providecommand \bibnamefont  [1]{#1}%
\providecommand \bibfnamefont [1]{#1}%
\providecommand \citenamefont [1]{#1}%
\providecommand \href@noop [0]{\@secondoftwo}%
\providecommand \href [0]{\begingroup \@sanitize@url \@href}%
\providecommand \@href[1]{\@@startlink{#1}\@@href}%
\providecommand \@@href[1]{\endgroup#1\@@endlink}%
\providecommand \@sanitize@url [0]{\catcode `\\12\catcode `\$12\catcode
  `\&12\catcode `\#12\catcode `\^12\catcode `\_12\catcode `\%12\relax}%
\providecommand \@@startlink[1]{}%
\providecommand \@@endlink[0]{}%
\providecommand \url  [0]{\begingroup\@sanitize@url \@url }%
\providecommand \@url [1]{\endgroup\@href {#1}{\urlprefix }}%
\providecommand \urlprefix  [0]{URL }%
\providecommand \Eprint [0]{\href }%
\providecommand \doibase [0]{http://dx.doi.org/}%
\providecommand \selectlanguage [0]{\@gobble}%
\providecommand \bibinfo  [0]{\@secondoftwo}%
\providecommand \bibfield  [0]{\@secondoftwo}%
\providecommand \translation [1]{[#1]}%
\providecommand \BibitemOpen [0]{}%
\providecommand \bibitemStop [0]{}%
\providecommand \bibitemNoStop [0]{.\EOS\space}%
\providecommand \EOS [0]{\spacefactor3000\relax}%
\providecommand \BibitemShut  [1]{\csname bibitem#1\endcsname}%
\let\auto@bib@innerbib\@empty
\end{thebibliography}%


\vspace{0mm}
\begin{thebibliography}{99}
\vspace{0mm}

\bibitem{Kibble1976} T.W. B. Kibble, Topology of cosmic domains, and strings,
J. Phys. A. {\bf 9}, 1387 (1976).
\bibitem{Kibble1980} T.W. B. Kibble, Some implications of a cosmological phase
transition, Phys. Rep. {\bf 67}, 183 (1980).
\bibitem{Zurek1985} W. H. Zurek, Cosmological experiments in superfluid
helium?, Nature (London) {\bf 317}, 505 (1985).
\bibitem{Zurek1996} W. H. Zurek, Cosmological experiments in condensed
matter systems, Phys. Rep. {\bf 276}, 177 (1996).
\bibitem{Polkovnikoc2011} A. Polkovnikov, K. Sengupta, A. Silva, and M. Vengalattore,
Colloquium: Nonequilibrium dynamics of closed interacting
quantum systems, Rev. Mod. Phys. {\bf 83}, 863 (2011).
\bibitem{Campo2014} A. Del Campo and W. Zurek, Universality of phase
transition dynamics: Topological defects from symmetry
breaking, Int. J. Mod. Phys. A {\bf 29}, 1430018 (2014).

\bibitem{Chuang1991} I. Chuang, B. Yurke, R. Durrer, and N. Turok, Cosmology in the Laboratory: Defect Dynamics in Liquid Crystals, Science {\bf 251},
1336 (1991).
\bibitem{Bowick1994} M. J. Bowick, L. Chandar, E. A. Schik, and A. M. Srivastava, The Cosmological Kibble Mechanism in the Laboratory: String Formation in Liquid Crystals, Science {\bf 263}, 943 (1994).
\bibitem{Digal1999} S. Digal, R. Ray, and A. M. Srivastava, Observing Correlated Production of Defects and Antidefects in Liquid Crystals, Phys. Rev. Lett. {\bf 83}, 5030 (1999).
\bibitem{Hendry1994} P. C. Hendry, N. S. Lawson, R. A. M. Lee, P. V. E.
McClintock, and C. D. H. Williams, Generation of defects
in superfluid $^4$He as an analogue of the formation of cosmic
strings, Nature (London) {\bf 368}, 315 (1994).
\bibitem{Bauerle1996} C. B$\ddot{a}$uerle, Y. M. Bunkov, S. N. Fisher, H. Godfrin, and
G. R. Pickett, Laboratory simulation of cosmic string formation
in the early Universe using superfluid 3He, Nature
(London) {\bf 382}, 332 (1996).
\bibitem{Ruutu1996} V. M. H. Ruutu, V. B. Eltsov, A. J. Gill, T.W. B. Kibble, M.
Krusius, Yu. G. Makhlin, B. Plaais, G. E. Volovik, and Wen
Xu, Vortex formation in neutron-irradiated superfluid $^3$He as
an analogue of cosmological defect formation, Nature
(London) {\bf 382}, 334 (1996).
\bibitem{Dodd1998} M. E. Dodd, P. C. Hendry, N. S. Lawson, P. V. E.
McClintock, and C. D. H. Williams, Nonappearance of
Vortices in Fast Mechanical Expansions of Liquid 4He
through the Lambda Transition, Phys. Rev. Lett. {\bf 81}, 3703
(1998).
\bibitem{Carmi2000} R. Carmi, E. Polturak, and G. Koren, Observation of Spontaneous Flux Generation in a Multi-Josephson-Junction Loop, Phys. Rev. Lett. {\bf 84}, 4966 (2000).
\bibitem{Monaco2002} R. Monaco, J. Mygind, and R. J. Rivers, Zurek-Kibble Domain Structures: The Dynamics of Spontaneous Vortex Formation in Annular Josephson Tunnel Junctions, Phys. Rev. Lett. {\bf 89}, 080603 (2002).
\bibitem{Monaco2003} R. Monaco, J. Mygind, and R. J. Rivers, Spontaneous fluxon formation in annular Josephson tunnel junctions, Phys. Rev. B { \bf 67},104506 (2003).
\bibitem{Monaco2006} R. Monaco, J. Mygind, M. Aaroe, R. J. Rivers, and V. P.
Koshelets, Zurek-Kibble Mechanism for the Spontaneous Vortex Formation in $\mathrm{Nb}\mathrm{\text{\ensuremath{-}}}\mathrm{Al}/{\mathrm{Al}}_{\mathrm{ox}}/\mathrm{Nb}$ Josephson Tunnel Junctions: New Theory and Experiment, Phys. Rev. Lett. {\bf 96}, 180604 (2006).
\bibitem{Maniv2003} A. Maniv, E. Polturak, and G. Koren, Observation of Magnetic Flux Generated Spontaneously During a Rapid Quench of Superconducting Films, Phys. Rev. Lett. {\bf 91}, 197001 (2003).
\bibitem{Golubchik2010} D. Golubchik, E. Polturak, and G. Koren, Evidence for Long-Range Correlations within Arrays of Spontaneously Created Magnetic Vortices in a Nb Thin-Film Superconductor, Phys. Rev. Lett. {\bf 104}, 247002 (2010).
\bibitem{Xu2014} X.-Y. Xu, Y.-J. Han, K. Sun, J.-S. Xu, J.-S. Tang, C.-F. Li,
and G.-C. Guo, Quantum Simulation of Landau-Zener Model Dynamics Supporting the Kibble-Zurek Mechanism, Phys. Rev. Lett. {\bf 112}, 035701 (2014).
\bibitem{Pyka2013} K. Pyka, J. Keller, H. L. Partner, R. Nigmatullin, T.
Burgermeister, D.M. Meier, K. Kuhlmann, A. Retzker,
M. B. Plenio, W. H. Zurek, A. del Campo, and T. E.
Mehlstubler, Topological defect formation and spontaneous
symmetry breaking in ion Coulomb crystals, Nat. Commun.
{ \bf 4}, 2291 (2013).
\bibitem{Ulm2013} S. Ulm, J. Ronagel, G. Jacob, C. Degunther, S. T. Dawkins,
U. G. Poschinger, R. Nigmatullin, A. Retzker, M. B. Plenio,
F. Schmidt Kaler, and K. Singer, Observation of the Kibble
Zurek scaling law for defect formation in ion crystals, Nat. Commun. {\bf 4}, 2290 (2013).
\bibitem{Ejtemaee2013} S. Ejtemaee and P. C. Haljan, Spontaneous nucleation and
dynamics of kink defects in zigzag arrays of trapped ions,
Phys. Rev. A. {\bf 87}, 051401(R) (2013).
\bibitem{Sadler2006} L. E. Sadler, J.M. Higbie, S. R. Leslie, M. Vengalattore, and
D. M. Stamper-Kurn, Spontaneous symmetry breaking in a
quenched ferromagnetic spinor Bose-Einstein condensate,
Nature (London) {\bf 443}, 312 (2006).
\bibitem{Weiler2008} C. N. Weiler, T.W. Neely, D. R. Scherer, A. S. Bradley,
M. J. Davis, and B. P. Anderson, Spontaneous vortices in the
formation of Bose-Einstein condensates, Nature (London)
{\bf 455}, 948 (2008).
\bibitem{Lamporesi2013} G. Lamporesi, S. Donadello, S. Serafini, F. Dalfovo, and G.
Ferrari, Spontaneous creation of Kibble-Zurek solitons in a
Bose-Einstein condensate, Nat. Phys. {\bf 9}, 656 (2013).
\bibitem{Navon2015} N. Navon, A. L. Gaunt, R. P. Smith, and Z. Hadzibabic,
Critical dynamics of spontaneous symmetry breaking in a
homogeneous Bose gas, Science {\bf 347}, 167 (2015).
\bibitem{Donadello2016} S. Donadello, S. Serafini, T. Bienaim, F. Dalfovo, G.
Lamporesi, and G. Ferrari, Creation and counting of defects
in a temperature-quenched Bose-Einstein condensate, Phys.
Rev. A {\bf 94}, 023628 (2016).
\bibitem{Ko2019} B. Ko, J.W. Park, and Y. Shin, Kibble-Zurek universality in
a strongly interacting Fermi superfluid, Nat. Phys. {\bf 15}, 1227
(2019).



 \bibitem{Campo2018} A. del Campo, Universal Statistics of Topological Defects Formed in a Quantum Phase Transition, Phys. Rev. Lett. {\bf 121}, 200601 (2018).
\bibitem{Ruiz2020} F. J. Gmez-Ruiz, J. J. Mayo, and A. del Campo, Full Counting Statistics of Topological Defects after Crossing a Phase Transition, Phys. Rev. Lett. {\bf 124}, 240602 (2020).
\bibitem{Campo2021}A. del Campo, F. J. Gomez-Ruiz, Z. Li, C. Xia, H. Zeng,
and H. Zhang, Universal statistics of vortices in a newborn
holographic superconductor: Beyond the Kibble-Zurek
mechanism, J. High Energy Phys.  06 (2021) 061.
\bibitem{Cui2020} J.-M. Cui, F. J. Gomez-Ruiz, Y.-F. Huang, C.-F. Li, G.- C.
Guo, and A. del Campo, Commun. Experimentally testing quantum critical dynamics beyond the Kibble-Zurek mechanism,  Commun. Phys. {\bf 3}, 44 (2020).

\bibitem{Goo2021} Junhong Goo, Younghoon Lim, and Y. Shin, Defect Saturation in a Rapidly Quenched Bose Gas,
Phys. Rev. Lett. {\bf 127}, 115701 (2021).
\bibitem{Campo2010}A. del Campo, G. De Chiara, G. Morigi, M. B. Plenio, and
A. Retzker, Structural Defects in Ion Chains by Quenching
the External Potential: The Inhomogeneous Kibble-Zurek
Mechanism, Phys. Rev. Lett. {\bf 105}, 075701 (2010).



\bibitem{Liu2018v1} I.-K. Liu, S. Donadello, G. Lamporesi, G. Ferrari, S.-C.
Gou, F. Dalfovo, and N. P. Proukakis, Dynamical equilibration
across a quenched phase transition in a trapped
quantum gas, Commun. Phys. {\bf 1}, 24 (2018).
\bibitem{Chesler2015} P. M. Chesler, A.M. Garcia-Garcia, and H. Liu, Defect
Formation beyond Kibble-Zurek Mechanism and Holography,
Phys. Rev. X {\bf 5}, 021015 (2015).
\bibitem{Corman2014} L. Corman, L. Chomaz, T. Bienaime, R. Desbuquois, C.
Weitenberg, S. Nascimbene, J. Dalibard, and J. Beugnon,
Phys. Rev. Lett. {\bf 113}, 135302 (2014).
\bibitem{Das2012} A. Das, J. Sabbatini, and W. H. Zurek, Winding up superfluid in a torus via Bose Einstein condensation, Sci. Rep. { \bf 2}, 352
(2012).
\bibitem{Sonner2015} J. Sonner, A. del Campo, and W. H. Zurek, Universal far-from-equilibrium dynamics of a holographic superconductor, Nat. Commun. {\bf 6}, 7406 (2015).
\bibitem{Xia2020} C. Y. Xia and H. B. Zeng, Winding up a finite size holographic superconducting ring beyond
Kibble-Zurek mechanism, Phys. Rev. D {\bf 102}, 12, 126005 (2020).
\bibitem{Maldacena1998} J. M. Maldacena, The Large-N Limit of Superconformal Field Theories and Supergravity, Adv. Theor. Math. Phys. {\bf 2}, 231 (1998).
\bibitem{Gubser1998} S. S. Gubser, I. R. Klebanov, and A. M. Polyakov, Gauge theory correlators from non-critical string theory, Phys.
Lett. B. {\bf 428}, 105 (1998).
\bibitem{Witten1998} E. Witten, Dynamical equilibration across a quenched phase transition in a trapped quantum gas, Adv. Theor. Math. Phys {\bf 2}, 253 (1998).

\bibitem{Zaanen:2015oix}
  J.~Zaanen, Y.~W.~Sun, Y.~Liu and K.~Schalm,
  ``Holographic Duality in Condensed Matter Physics,''
  Cambridge University Press, (2015).
\bibitem{Ammon2015} M. Ammon and J. Erdmenger, \Gauge/gravity duality:
Foundations and applications," Cambridge University
Press, 2015.
\bibitem{Liu2018} H. Liu and J. Sonner, Holographic systems far from equilibrium: a review, arXiv:1810.02367.
\bibitem{Jan2021}  J. Zaanen, Lectures on quantum supreme matter, arXiv:2110.00961.

\bibitem{Gubser2008} S. S. Gubser, Breaking an Abelian gauge symmetry near a black hole horizon, Phys. Rev. D {\bf 78}, 065034 (2008).
\bibitem{Hartnoll2008} S. A. Hartnoll, C. P. Herzog, and G. T. Horowitz, Building a Holographic Superconductor, Phys. Rev. Lett. { \bf 101}, 031601 (2008).
\bibitem{Zeng2021} H. B. Zeng, C. Y. Xia, and H. Q. Zhang, Topological defects as relics of
spontaneous symmetry breaking in a holographic superconductor, J. High Energ. Phys. 03 (2021) 136 .




\end{thebibliography}
\end{document}